\documentclass{ws-ijmpd}
\usepackage[super,compress]{cite}
\usepackage{color}
\begin{document}

\markboth{Ebisuzaki et al.}{INO: Interplanetary Network of Optical Lattice Clocks}

%
%
\title{INO: Interplanetary Network of Optical Lattice Clocks}

\author{TOSHIKAZU EBISUZAKI}
\address{Institute of Physical \& Chemical Research (RIKEN), Hirosawa, 
Wako, Saitama 351-0198, Japan\\
ebisu@riken.jp}

\author{HIDETOSHI KATORI}
\address{Institute of Physical \& Chemical Research (RIKEN), Hirosawa, 
Wako, Saitama 351-0198, Japan \\
Graduate School of Engineering, Department of Applied Physics, 
The University of Tokyo, Hongo, 
Bunkyo, Tokyo 113-8656, Japan
\\
hkatori@riken.jp}

\author{JUN'ICHIRO MAKINO}
\address{Institute of Physical \& Chemical Research (RIKEN), Hirosawa, 
Wako, Saitama, 351-0198 Japan \\
Graduate School of Science, Kobe University, Rokkodai, 
Nada, Kobe, Hyogo 657-8501, Japan
\\
makino@mail.jmlab.jp}

\author{ATSUSHI NODA}
\address{Japan Aerospace Exploration Agency (JAXA), Sengen, Tsukuba, Ibaraki 305-8505, Japan\\
noda.atsushi@jaxa.jp}

\author{HISAAKI SHINKAI}
\address{Osaka Institute of Technology, Kitayama, Hirakata, Osaka 573-0196, Japan\\
Institute of Physical \& Chemical Research (RIKEN), Hirosawa, Wako, Saitama, 351-0198 Japan \\
hisaaki.shinkai@oit.ac.jp}

\author{TORU TAMAGAWA}
\address{Institute of Physical \& Chemical Research (RIKEN), Hirosawa, 
Wako, Saitama 351-0198, Japan \\
tamagawa@riken.jp}

\maketitle


\begin{abstract}
The new technique of measuring frequency by optical lattice clocks now approaches to the relative precision of $(\Delta f/f)=O(10^{-18})$. 
We propose to place such precise clocks in space and to use Doppler tracking method for detecting low-frequency gravitational wave below 1 Hz.  
Our idea is to locate three spacecrafts at one A.U. distance (say at L1, L4 \& L5 of the Sun-Earth orbit), and apply the Doppler tracking method by communicating ``the time" each other. 
Applying the current available technologies, we obtain the sensitivity for gravitational wave with three or four-order improvement ($h_{\rm n}\sim 10^{-17}$ or $10^{-18}$ level in $10^{-5}$Hz -- $1$ Hz) than that of Cassini spacecraft in 2001. 
This sensitivity enables us to observe black-hole mergers of their mass greater than $10^5 M_\odot$ in the cosmological scale. 
Based on the hierarchical growth model of black-holes in galaxies, we estimate the event rate of detection will be 20-50 a year. 
We nickname "INO" (Interplanetary Network of Optical Lattice Clocks) for this system, named after Tadataka Ino (1745--1818), a Japanese astronomer, cartographer, and geodesist. 
\end{abstract}


\ccode{PACS numbers: 04.80.Nn, 95.55.Ym, 95.85.Sz, 06.30.Ft, 95.55.Sh}


\section{Introduction}
The direct detections of gravitational wave (GW) by LIGO/Virgo groups\cite{GW150914PRL} has opened new era for physics and astronomy. 
With this new method of observing the Universe, we are now able to observe black-holes (BHs) directly with GW. 
With help of other observational bands\cite{GW170817PRL}, {\it i.e.} with the electromagnetic waves from radio to 
gamma-ray or by neutrino observation, we can discuss the details of high-energy events, the equation of state of nuclear matter, 
cosmology, and the validity of gravitational theories.
LIGO/Virgo groups announced so-far that ten events of the coalescences of binary BHs
\cite{GW150914PRL,GW151226PRL,GW170104PRL,GW170608ApJ,GW170814PRL,GWTC1}, 
and one merger of binary neutron stars\cite{GW170817PRL}. 
In 2019, LIGO/Virgo detecters will start their observation again with upgraded systems, and KAGRA in Japan will start 
its observation, too.  Such GW observations will give us the statistics of the events and make the science more 
precise and trustable.

Among the unsolved problems in the Universe, however, the growth process of large BHs is left untouched. 
Almost all of the galaxies in the Universe have super-massive black holes (SMBHs) in their center, whose mass is over $10^6 M_\odot$ \cite{Kormendy:2013db}. 
Observational data show that the masses of such SMBHs are proportional to the masses of their bulge (the center part of the galaxies) \cite{Magorrian:1998cs, Ferrarese:2000je}.  
This fact indicates that the SMBHs in the galaxies coevolved with its mother galaxy, but such an inevitable relation is still a mystery in the history of the Universe.  

One of the plausible scenario is the hierarchical growth model of the stars in the galaxy. 
This model says that a black hole (a seed black hole) is first formed in the center of galaxy when it grows to the size of a dwarf galaxy. 
The intermediate-mass black holes (IMBHs), whose mass is around 100$\sim$1000~M$_{\odot}$, in its star clusters will then accumulate in the center region of the galaxy, merge together, and form to a SMBH \cite{Kawabe:2001wx}.  
This theory naturally explains central BH-bulge mass relation, described above.  
A seed black hole might be formed by another process; one possible way is to start from forming a $10^5 M_\odot$ black hole in the early stage of the Universe.  
Hierarchical mergers will proceed between galaxies, and such processes will naturally produce SMBHs \cite{Makino:2004ka, Matsubayashi:2007dx}.  
In result, the mergers of $10^6$ -- $10^{10} M_\odot$ black holes are expected to be the sources of GW\cite{Matsubayashi:2004it, Shinkai:2017gb}. 

Formation of SMBHs are also modeled by accumulations of large amount of gas to a seed BH \cite{Pezzulli:2016fc}.  Therefore, collecting the detections of GWs from large mass BHs, which means the direct evidence of the growth process of the BHs, is the clue to solve the current mystery of coevolution of a BH and its mother galaxy. 

All the gravitational detectors which are under operation today, such as LIGO and Virgo (and coming KAGRA\cite{KAGRAnatureastro2019}), are located on the ground, which means that we are hard to detect GW below 10~Hz since seismic vibration dominates as noise. 
The mergers of BHs in the range  $10^4$--$10^8 M_\odot$, on the other hand, produce GW 
in the so-called low-frequency band \cite{KurodaNiPan}, and end up 
below 1~Hz (See Fig.\ref{fig:char_strain} in this article). 
There are many proposals to detect GWs in the low- (including mid-) frequency band \cite{eLISA,AIGSO,AMIGO1,AMIGO2,MIGA,SOGRO,SOGRO2,ZAIGA}. 
Among them, we think that placing detectors in space (see a review by Ni \cite{WeiTouNi2016}) is the best and only possible way to detect GW signal from IMBHs. 

The project of ``evolved Laser Interferometer Space Antenna" (eLISA) by ESA \cite{eLISA} is the plan of constructing laser interferometer in space with the arm length $1.0 \times 10^6$ km, targeting mainly at milli-Hz range of GW. 
Locating three spacecrafts at Earth-like solar orbits with $10$-degree lag
with drag-free flight motion, and using the light-transponder technique, ESA plans to realize the system in early 2030s.  

Japanese group proposed ``DECi-hertz Interferometer GW Observatory" (DECIGO/B-DECIGO) project \cite{Nakamura:2016cn}, which plans to construct 
Fabry-Perot laser interferometer with 1000~km (100~km) arm length, with three spacecrafts on the Sun-Earth orbit (around the Earth orbit) with drag-free flight motion.  Their main target is deci-Hz range of GW. 

Space-borne interferometers such as eLISA or B-DECIGO require significant technical breakthroughs.  
We, in the present article, propose an alternative method for detecting low-frequency GWs, technically feasible with the current technologies.  
Our idea is to locate three spacecrafts at A.U. scales (say at L1, L4 and L5 of the Sun-Earth orbit), which load 
the optical lattice clock, the ultimate precise atomic clock. 
By comparing the time each other, applying the principle of the Doppler tracking, we can detect the passage of GWs of mHz range. We named this proposal {\bf\it ``Interplanetary Network of Optical Lattice Clocks"}, with the abbreviation {\bf\it INO}. 
The acronym, INO, is named after Tadataka Ino, a Japanese astronomer, cartographer, and geodesist, who made precise map of Japan two centuries ago. 

The ideas of locating spacecrafts at A.U. scales have already been appeared \cite{AMIGO1,SuperASTROD}. 
For example, Ni introduced a mission concept of ``Astrodynamical Space Test of Relativity using Optical Devices" (ASTROD) \cite{1212.2816,AMIGO1} and Super-ASTROD\cite{SuperASTROD}.  ASTROD is to locate spacecrafts at Sun-Earth L3/L4/L5, while Super-ASTROD 
is to locate spacecrafts at Sun-Jupiter L3/L4/L5 together with at Sun-Earth L1/L2, and to probe primordial GWs at $10^{-6}$--$10^{-3}$ Hz by Doppler tracking using laser pulse ranging and precision optical clocks \cite{WeiTouNi2016}.

The ideas of using atomic clocks for detecting GW have also already been appeared \cite{Tinto2009, 1501.00996, Vutha2015, Kolkowitz:2016gx,SuWangWangJetzer}.  
For example, 
Su et al. \cite{SuWangWangJetzer} proposed a similar idea naming ``Double Optical Clocks in Space" (DOCS). They proposed to place two optical lattice clocks at the Lagrange points of the Earth-Moon orbit, and link them with the Earth by radio. 

We, in this article, discuss more feasibilities and technological new idea together with detectable distance of the detectors and GW sources counts.  Our proposal does not reach to the best sensitivities and detectable distance than so-far proposed concepts.  However, we shall show that our concept is enough for testing a SMBH-formation scenario 
with the currently available technologies within a certain operation period.  

\section{GW detection using optical lattice clocks}
The only observations of GW in space so far are the one by tracking artificial spacecraft using Doppler effect \cite{BraginskyGertsenshtein,Thorne:1976tl} (actually they showed us the upper-bound constraint of the GW). 
The method is to observe the velocity shift (Doppler shift) produced by passing GW between the Earth and the spacecraft, by comparing the frequency of the signal sent from the Earth and received at the spacecraft using their clocks. 
The sensitivity of this Doppler-tracking method depends on the distance of the signal baseline. 
Until now, the most strict sensitivity was obtained by the Cassini spacecraft which was launched for surveying the Saturn
\cite{Armstrong:2003}.  

Armstrong \cite{Armstrong:2006hw} listed up key noises to be improved in the future mission of Cassini-type GW observation; (i) frequency standard, (ii) ground electronics, (iii) tropospheric scintillation, (iv) plasma scintillation, (v) spacecraft motion, 
and (iv) antenna mechanical [See Table \ref{table:armstrong}]. 
We discuss how we can improve them with the current technologies one by one. 

\begin{table}[h]
\tbl{ Required improvement in subsystems to improve overall Doppler sensitivity by a factor of 10 relative to Cassini-era performance. (Copy of Table 4 of Armstrong (2006) \cite{Armstrong:2006hw}, added the left column.)
\label{table:armstrong}}
{\begin{tabular}{clp{6cm}l} \toprule
&Noise source & Comment ($\sigma_y$ at $\tau=1000$s) & Required \\
& & current values &improvement \\
\colrule
(i)& Frequency standard & FTS $+$ distribution $\simeq 8 \times 10^{-16}$&  $\simeq 8$X \\
(ii) &Ground electronics& $\simeq 2 \times 10^{-16}$ & $\simeq 2$X\\
(iii)&Tropospheric scintillation& $\simeq 10^{-15}$ under favorable conditions& $\simeq 10$X\\
(iv) &Plasma scintillation&Cassini-class radio system probably adequate for calibration to $\simeq 10^{-16}$& $\simeq 1$X\\
(v) &Spacecraft motion &$\simeq 2 \times 10^{-16}$ & $\simeq 2$X\\
(iv) &Antenna mechanical&$\simeq 2 \times 10^{-15}$ under favorable conditions & $\simeq 20$X\\
\botrule
\end{tabular}
}
\end{table}

\subsection{Usage of optical lattice clocks}
The key technology of the Doppler-tracking method is the stability of clocks [the list (i)].
We propose to use ``optical lattice clocks" that allow significant improvement in atomic clocks’ stability.  Atomic clocks steer the frequency of local oscillators, such as cavity-stabilized lasers, by referencing atomic transitions.  Stability of such atomic clocks are limited by the quantum projection noise \cite{Itano1993} that is given by the number of atoms $N$.  By interrogating $N\sim 10^6$ atoms trapped at the anti-nodes of a standing-wave laser, which is referred to as an optical lattice, and by eliminating the Stark shift perturbation by tuning the laser to the magic frequency \cite{Katori2003,Takamoto2003}, the optical lattice clocks achieve \cite{Takano:es} high stability and accuracy approaching $10^{-18}$.  By applying an operational magic frequency, the accuracy of $10^{-19}$ is in scope \cite{Katori2015}.  If this level of the accuracy is obtained, then the noise from clock can be completely ignored.

Kolkowitz {\it et al} \cite{Kolkowitz:2016gx} proposed to use optical lattice clocks to detect GWs.  Their idea is to measure the Doppler shift between the two optical lattice clocks in space which are communicating with lasers.  
By controlling two mirrors located apart in the drag-free state, they propose to measure the frequency difference between two optical lattice clocks using precise laser which are linked with these mirrors.  The core idea is the same with that of the Doppler-tracking method, but with the technology of drag-free control, they say the sensitivity is greatly improved at 0.01$\sim$1~Hz.  
However, there is a disadvantage in drag-free technology.  The remained acceleration of free mass is controlled with magnetic field, but 
when cosmic-ray hits the device, the photoelectric effect charges the free-mass and this fluctuation behaves noise.  Especially at the lower frequency, the residual error is inversely proportional to the square of frequency\cite{Armano:2017ij}, and as a result the sensitivity of their proposal is at the same level with eLISA. 

We therefore propose not to use drag-free control, but to improve Doppler-tracking method with advanced optical lattice clock and the light-linking technology for constructing a GW detector. 

Ni\cite{1212.2816,WeiTouNi2016} mentioned that it is important to separate perturbations of solar-system bodies with GW signals.  We believe that as far as we seek GW from merging IMBHs, such perturbations by planets or other small bodies are distinguishable by the Fourier spectrum of the motions of sources. 

\subsection{Location of INO spacecrafts}
If we measure all the difference of the clock between the spacecrafts in space, then we do not need to care  
the noises due to ground electronics [the list (ii)] and tropospheric scintillation [the list (iii)]. 

\begin{figure}[h]
\begin{center}
\includegraphics[width=0.75\linewidth]{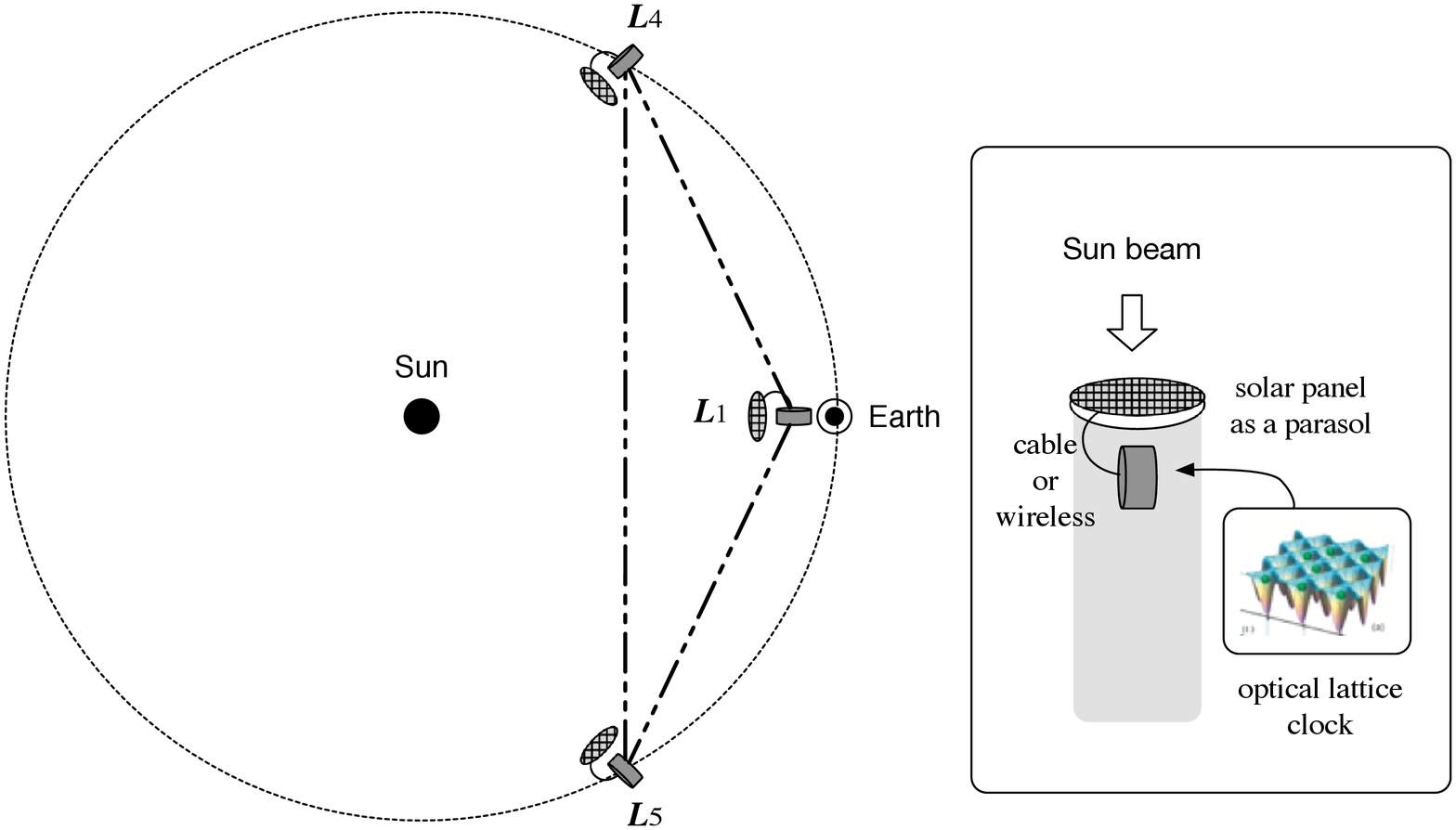}
\end{center}
\caption{A planned location of the spacecrafts: Lagrangian points $L_1$, $L_4$ and $L_5$ of the Sun-Earth orbit. The $L_1$ is at 
1/100 A.U. from the Earth, while $L_4$ and $L_5$ forms equilateral triangle with the Sun and the Earth respectively; the distance between $L_1$--$L_4$($L_5$) is 1 A.U., while that of $L_4$--$L_5$ is $\sqrt{3}$ A.U.
Two-frequency radio or light will be used for communication between spacecrafts.  The inset explains that the solar panel of the  spacecrafts is separated as a parasol from the main body, in order to prevent acceleration noise due to solar wind. 
\label{fig:satellite_config}}
\end{figure}

As we already mentioned, it is preferable to locate the spacecrafts for Doppler tracking at far distance such as beyond the orbit of Jupiter or Saturn. This request, however, is severe for keeping power and fuel. 
We therefore propose to locate three spacecrafts at the L1, L4, and L5 of the Sun-Earth orbit, which enable us to take the baselines between each spacecrafts at the order of A.U.  (see Fig.\ref{fig:satellite_config}). 

\subsection{Communication between the spacecrafts}
The noise list (iv) by Armstrong is plasma scintillation in the Solar system. 
We may have two ways to communicate each spacecraft; radio or light.  
 If we link them with light, the phase fluctuation by the plasma effects is negligible. However, light communication requires precise directivity than radio.  In the current technology, for communication over $10^7$ km, radio is preferable.  If we link them with radio, then double tracking method which uses two-frequency bands will compensate the phase shift due to interplanetary plasma.

\subsection{Separation of the main body and the solar-cell panel}
The noise list (v) by Armstrong is spacecraft stability against radiation pressure of the Sun beam, which dominates the noise in the lower frequency range.  Suppose we require the sensitivity of $h_{\rm n}=10^{-17}$ for the baseline of $1.5 \times 10^8$ km ($\sim$ 1 A.U.), which is comparable with the amplitude 0.7~mm; that corresponds to the 
measurement of the velocity 2$\times$10$^{-8}$~m~s$^{-1}$, or the acceleration 1.4$\times$10$^{-12}$~m~s$^{-2}$
at 10$^{-5}$~Hz. 

The radiation pressure force is $F=P/c$, where $P$ is the power of the Sun beam and $c$ is the speed of light. Around the Earth orbit, $P$ is 1.3~kW m$^{-2}$ per unit area.  
If we suppose 
the solar cell panel as 10~m$^2$ and the mass of the spacecraft as 1000~kg, then the acceleration of the spacecraft due to the beam pressure is about 5$\times$10$^{-8}$ m s$^{-2}$. 
The pressure of the Sun fluctuates at the order of $10^{-3}$, so that the acceleration fluctuates at the order of 10$^{-11}$.  
In order to detect the GW of which acceleration is at the order of 10$^{-12}$, 
we should reduce the fluctuation one order smaller.  
This is attainable by mechanically separating the spacecraft main body from its solar-cell panel and use the solar panel as a parasol for shielding from the Sun beam (see the inset of Fig.\ref{fig:satellite_config}).  
The area of the solar-cell is around  270 W/$m^2$, effective than the normal spacecraft since this network always receive the Sun beam.  
If the cable connection is a source of vibration, then the wireless 
transmission of electricity can be used.  In the current technology, 
wireless transmission of 12 kW with 10 cm is obtained at experiments on the ground.
This structure removes the dynamical interactions which may reduce the acceleration noise two-orders of magnitude. 

\section{Sensitivity of INO}
We estimate the reachable sensitivity for GW detection with current known technologies.  In order to make the most feasible discussion, we do not consider to use drag-free control, nor precise laser control, but simply apply the advanced optical lattice clock to the Doppler-tracking method. 

The sensitivity of the Doppler-tracking method is well understood by the report of Cassini spacecraft 
\cite{Armstrong:2006hw,Armano:2017ij}, which keeps the best record as 
$h_{\rm n}\sim3\times10^{-15}$ at 10$^{-4}$~Hz, where 
$h_{\rm n}$ is the noise amplitude, which is given by the square root of the combination of 
the power spectrum of the noise times frequency $f$. 
The noise amplitude is the standard quantity since it can be compared directly with the 
characteristic strain $h_{\rm c}$ which expresses the strength of the GW signal. 
Cassini's sensitivity showed the curve $f^{-1}$ below 10$^{-4}$~Hz.  

The origins of noise in Cassini are identified mostly from the accuracy of the atomic clock and from the fluctuation of troposphere of the Earth \cite{Armstrong:2006hw}. 
As we discussed in the previous section, 
if we use the advanced optical lattice clock instead of the atomic clock, and let the spacecrafts communicate
each other directly, and with a Sun-beam shield, 
the sensitivity will be dramatically improved. 
From the Table.4 in \cite{Armstrong:2006hw}, we estimate that 
the three or four-order improved version of Cassini spacecraft (i.e. the minimum sensitivity is around 
$h_{\rm n}=10^{-17}$ or $10^{-18}$) will be available.   

\begin{figure}[bh]
\begin{center}
\includegraphics[width=0.75\linewidth]{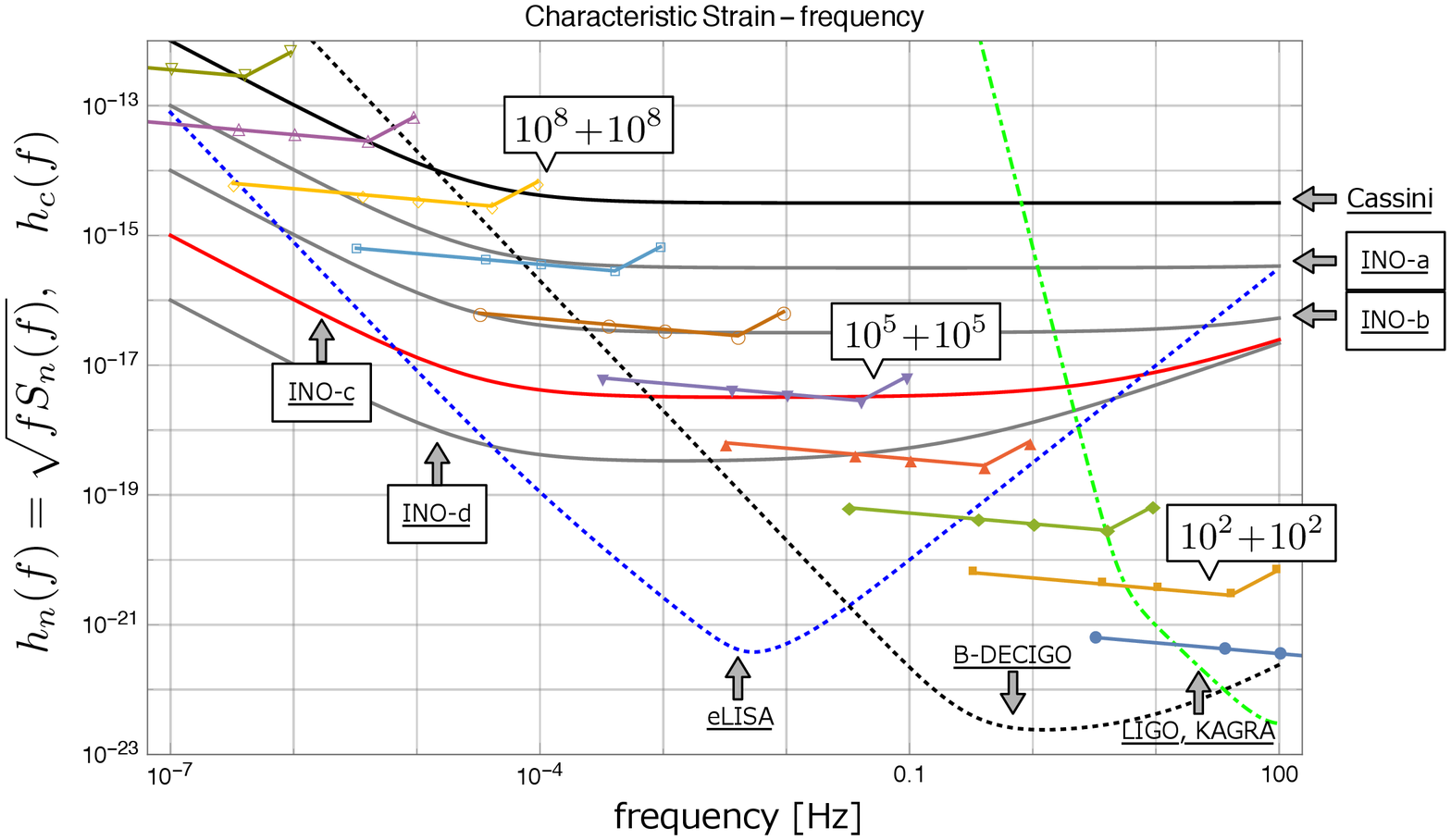}
\end{center}
\caption{
Sensitivity of Doppler-tracking spacecrafts and expected strains of GW.
The most upper solid curves indicates the sensitivity of Cassini spacecraft (2001), while the other solid curves are 
those of 1$\sim$4-order improved version (we named INO-a, INO-b, INO-c and INO-d, respectively). 
The dotted line is the sensitivity curve of eLISA. Almost horizontal lines with symbols indicate the characteristic strain of GW from a merger of equal-mass binary BHs at 1~Gpc. 
Each line is for the inspiral phase; starts from its separation 50 times of the event horizon radius, and ends at their merger  (frequency moves up higher for smaller separation). 
\label{fig:char_strain}}
\end{figure}

Fig. \ref{fig:char_strain} shows the sensitivity curve of Cassini spacecraft and their one to four-order improved version (we named INO-a, INO-b, INO-c and INO-d, respectively), together with that of eLISA, B-DECIGO and advanced LIGO/KAGRA.  
Since the frequency dependence at the lower frequency is different, INO achieves the same sensitivity with eLISA at 10$^{-5}$~Hz, and better than eLISA in the range less than that.

In Fig. \ref{fig:char_strain}, we also plot the characteristic strain of the GW ($h_{\rm c}$) from a merger of the binary BHs with its distance 1~Gpc from the Earth. 
We plotted for mergers of equal-mass BHs for several different masses.  Each line starts from its frequency when the binary's separation is 50 times of their event horizon radius, and ends at the frequency when they merge. 

We see that the mergers of SMBHs of 10$^7\sim$10$^8 M_{\odot}$ produce GW around 10$^{-4}$~Hz, which is detectable with INO at the signal-to-noise ratio (SNR) 10.

\section{Expected GW events}
\subsection{Detectable distance for BH mergers}
\begin{figure}[h]
\begin{center}
\includegraphics[width=0.75\linewidth]{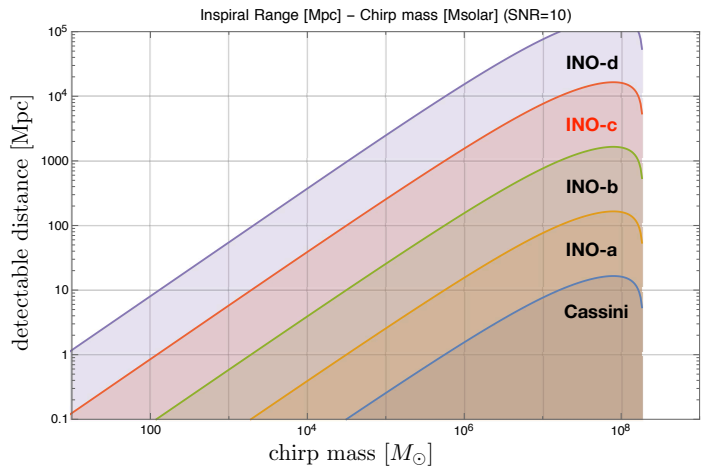}
\end{center}
\caption{
Detectable distance (observational distance, or horizon of the detector) of Cassini, INO-a, $\cdots$, INO-d as a function of the binary's chirp mass. The distance is the luminosity distance. All lines are for SNR=10. 
\label{fig:event_horizon}}
\end{figure}

Once the detector's sensitivity is given, then we can calculate the detectable distance (observational distance, or horizon of the detector) for typical BH merger events. 
In Fig. \ref{fig:event_horizon}, we plot them for Cassini and its improved version. We see INO-c  already covers the Universe for the mergers of SMBHs of their chirp mass 10$^7\sim$10$^8 M_{\odot}$.  
(For the binary of masses $m_1$ and $m_2$, the chirp mass, $M_c$, is given by  $M_c=(m_1m_2)^{3/5}/(m_1+m_2)^{1/5}$. $M_c$ determines the leading-order amplitude and frequency evolution of the gravitational-wave signal from inspiral binary.)

\subsection{Event rate of BH mergers by hierarchical growth model}
If we further assume the distribution model of BH mass (i.e. the evolution model of BHs), and
the distribution model of galaxies, together with cosmological model, then we can estimate the event rate per year.  

We calculate the event rate based on the hierarchical growth model \cite{Shinkai:2017gb}.  
This model assumes that the formation of SMBHs are from mergers of BHs in a hierarchical sequence. 
The number of BH mergers are estimated from the giant molecular cloud model, of its total numbers are depend on the 
size of galaxies.  The distribution and the size of galaxies are modeled from  
the number density of galaxies from the halo formation model.  
Although there are several unknown factors in the model (such as contributions of BH spins, mass ratio of binaries, 
merger ratio as a function of mass, etc), but the simplest model predicts that detection profile of the ground-based GW interferometers has a peak at 60 $M_\odot$, which is actually the same with the first detection, GW150914. 

\begin{figure}[h]
\begin{center}
\includegraphics[width=0.45\linewidth]{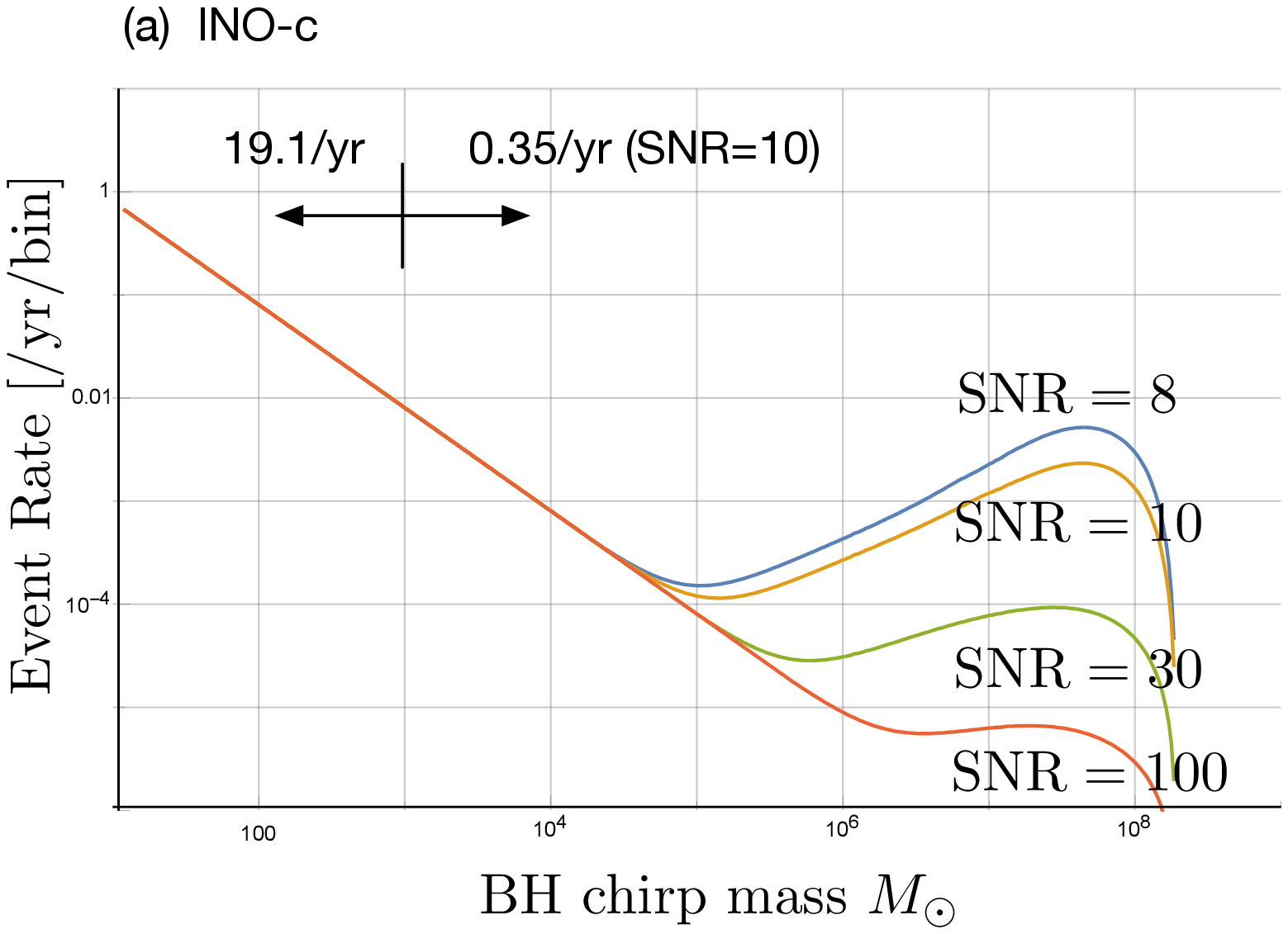}
\includegraphics[width=0.45\linewidth]{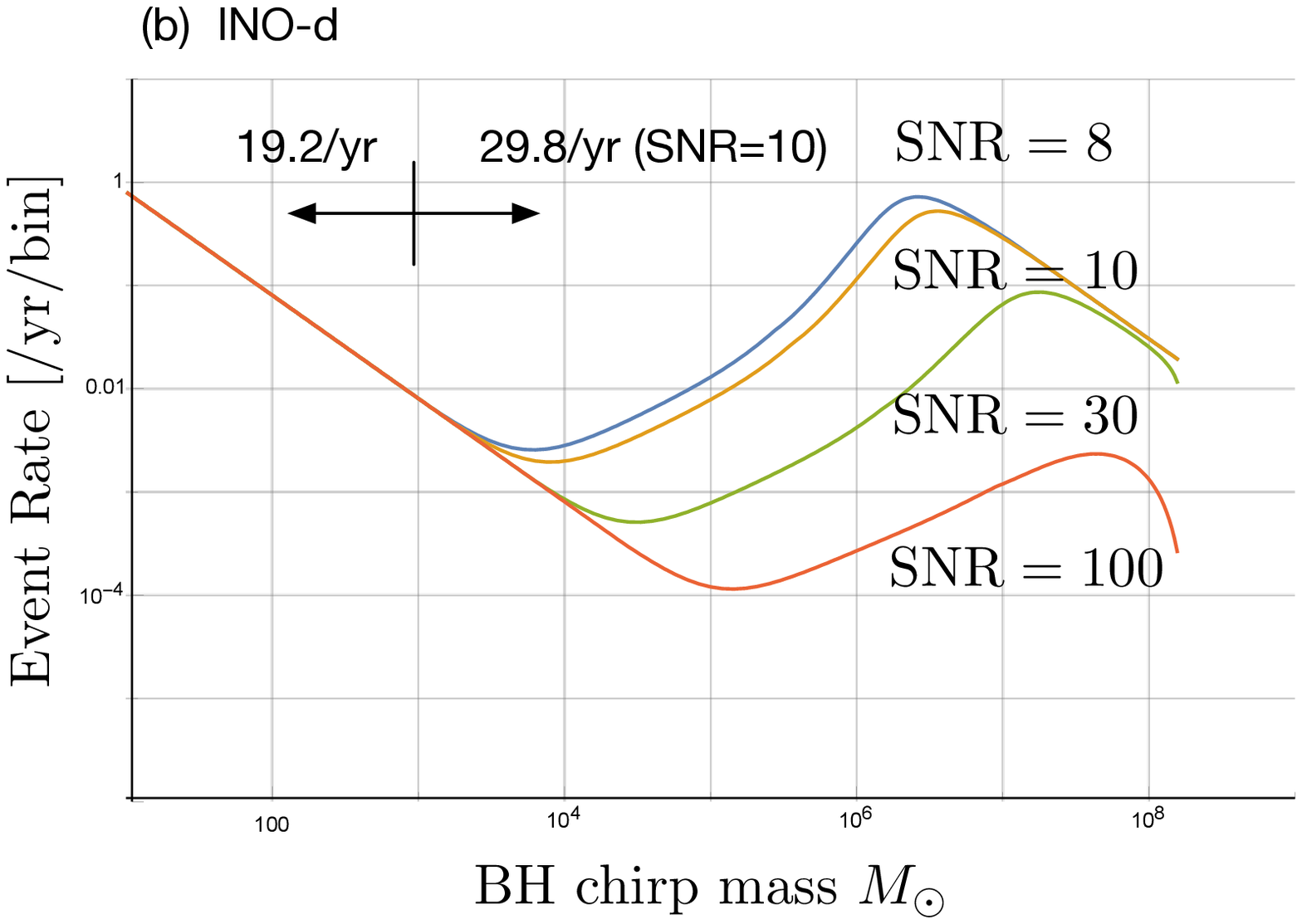}
\end{center}
\caption{Event rates of mergers of BHs by INO-c (a) and INO-d (b). 
Event rates are per year per bin, which we plot the number with 20 bins in log-scale in one order of the chirp mass. 
Each plot has four lines, for signal-to-noise ratio (SNR) of 8, 10, 30 and 100, respectively.  
The integrated event rates are also shown in the figure for upper and lower than $10^3M_\odot$. 
\label{fig:eventrate}}
\end{figure}

The result of the event rate for INO is shown in Figure \ref{fig:eventrate}. 
We show both for INO-c and INO-d.
The number is of per year per bin. If we integrated for the standard SNR=10 case, then we get 19.1 mergers for $<10^3
M_\odot$ and 0.35 mergers for $>10^3 M_\odot$ per year for INO-c, while we get 19.2 and 29.8 per year respectively for INO-d.


\section{Closing comments}

We proposed a new method for detecting GW in space, named INO (Interplanetary Network of Optical Lattice Clocks).  We discussed that,  with the current technologies, Cassini's Doppler tracking method (2001-2002) can be improved 3 to 4-order magnitudes.  Although even at INO-d level the best sensitivity is around $h_{\rm n}\sim 10^{-18}$, which is worse than the that of ongoing eLISA project, but we showed that INO-c and INO-d (which are of three and four-order improved sensitivity than Cassini, respectively) has better sensitivity range than eLISA at lower frequency range.  
We also showed that INO covers cosmological scale for observing BH mergers larger than $10^5 M_\odot$. 
We calculated event rates based on a hierarchical growth model of SMBHs, which say we could observe 
stellar-mass BH mergers 20 events per year at INO-c and if we can reach one order sensitivity up then 
we could observe more 30 BH mergers above $10^3 M_\odot$ range.  We think this number is worth trying to consider seriously. 

The detection of GW in space will give us the first clue to the process of the formation of SMBHs which is totally unknown now.  Our proposal is complementary to the method of interferometers, and the ultimate application of the optical lattice clocks.

\section*{Acknowledgment}
HS was supported by JSPS KAKENHI Grant Number JP17H06358 and JP18K03630.

\end{document}